%% file: 26DAFx_HardSync.tex
\def\papertitle{Alias-Free Oscillator Synchronization via Additive Synthesis}
\def\paperauthorA{Jonas Roth}
\def\paperauthorB{Domenic Keller}
\def\paperauthorC{Oscar Casta\~neda}
\def\paperauthorD{Christoph Studer}

\documentclass[twoside,a4paper]{article}
\usepackage{etoolbox}

\usepackage[print]{dafx26v3}

\usepackage{amsmath,amssymb,amsfonts,amsthm}
\usepackage{siunitx}
\usepackage{euscript}
\usepackage[T1]{fontenc}
\usepackage[utf8]{inputenc}
\usepackage{ifpdf}
\usepackage[english]{babel}
\usepackage{caption}
\usepackage{subcaption}
\usepackage{color}
\usepackage{booktabs}
\usepackage{lipsum}

\input glyphtounicode
\ninept

\input{macros/vmr-symbols-vecbold}
\input{macros/standard-macros}

\input{macros/defs}
\usepackage{siunitx}
\usepackage{microtype}
\usepackage{cuted}

\makeatletter
\newcommand{\ifblindelse}[2]{\if@blind #1\else #2\fi}
\makeatother

\newcommand{\acfollow}[1]{{#1}}
\newcommand{\acrot}[1]{\mathring{#1}}
\newcommand{\achard}[1]{\bar{#1}}
\newcommand{\acmirror}[1]{\hat{#1}}
\newcommand{\acpulsar}[1]{\ddot{#1}}

\newcommand{\flead}{\ensuremath{f_\text{lead}}\xspace}
\newcommand{\ffollow}{\ensuremath{f_\text{follow}}\xspace}
\newcommand{\fs}{f_\text{s}}
\newcommand{\Tlead}{\ensuremath{T_\text{lead}}\xspace}
\newcommand{\Tfollow}{\ensuremath{T_\text{follow}}\xspace}
\newcommand{\versinc}{\mathrm{versinc}}
\newcommand{\SINAD}{\ensuremath{\textit{SINAD}}\xspace}
\newcommand{\erm}{\mathrm{e}}
\newcommand{\jrm}{\mathrm{j}}

\newcounter{numauth}
\newcounter{listcnt}
\newcommand\authcnt[1]{\ifdefined#1 \stepcounter{numauth} \fi}

\newcommand\addauth[1]{
\ifdefined#1
\stepcounter{listcnt}
\ifnum \value{listcnt}<\value{numauth}
\appto\authorslist{, #1}
\else
\appto\authorslist{~and~#1}
\fi
\fi}
\authcnt{\paperauthorB}
\authcnt{\paperauthorC}
\authcnt{\paperauthorD}
\authcnt{\paperauthorE}
\authcnt{\paperauthorF}
\authcnt{\paperauthorG}
\authcnt{\paperauthorH}
\authcnt{\paperauthorI}
\authcnt{\paperauthorJ}
\def\authorslist{\paperauthorA}
\addauth{\paperauthorB}
\addauth{\paperauthorC}
\addauth{\paperauthorD}
\addauth{\paperauthorE}
\addauth{\paperauthorF}
\addauth{\paperauthorG}
\addauth{\paperauthorH}
\addauth{\paperauthorI}
\addauth{\paperauthorJ}

\usepackage{times}

\newif\ifpdf
\ifx\pdfoutput\relax
\else
   \ifcase\pdfoutput
      \pdffalse
   \else
      \pdftrue
   \fi
\fi

\ifpdf 
  \usepackage[pdftex,
    pdftitle={\papertitle},
    pdfauthor={\authorslist},
    pdfsubject={Proceedings of the 29th International Conference on Digital Audio Effects (DAFx26)},
    colorlinks=false, 
    bookmarksnumbered, 
    pdfstartview=XYZ 
  ]{hyperref}
  \usepackage[pdftex]{graphicx}
\else 
  \usepackage[dvips]{epsfig,graphicx}
  \usepackage[dvips,
    pdftitle={\papertitle},
    pdfauthor={\authorslist},
    pdfsubject={Proceedings of the 29th International Conference on Digital Audio Effects (DAFx26)},
    colorlinks=false, 
    bookmarksnumbered, 
    pdfstartview=XYZ 
  ]{hyperref}
\fi
\usepackage[hypcap=true]{caption}
\title{\papertitle}

\affiliation
{\paperauthorA, \paperauthorB, \paperauthorC, and \paperauthorD}
{Department of Information Technology and Electrical Engineering, ETH Zurich, Switzerland\\
{\tt \href{mailto:joroth@ethz.ch}{joroth@ethz.ch} | \href{mailto:domkeller@ethz.ch}{domkeller@ethz.ch} |
\href{mailto:caoscar@ethz.ch}{caoscar@ethz.ch} |
\href{mailto:studer@ethz.ch}{studer@ethz.ch}}
}

\usepackage{standalone}
\usepackage{tikz}
\usepackage{pgfplots}
\pgfplotsset{compat=1.18}

\begin{document}
\ifpdf 
  \DeclareGraphicsExtensions{.png,.jpg,.pdf}
\else  
  \DeclareGraphicsExtensions{.eps}
\fi


\maketitle

\sloppy

\begin{abstract}
Oscillator synchronization is a widely used sound-synthesis technique, but straightforward digital implementations suffer from aliasing artifacts. This paper presents an alias-free method for digital emulation of oscillator synchronization of arbitrary periodic waveforms based on additive synthesis. Starting from a finite set of Fourier-series coefficients representing a bandlimited free-running waveform, we derive linear spectral-resampling transforms that map these coefficients to those of the bandlimited synchronized waveform. Beyond conventional hard synchronization, the proposed approach also supports two additional soft-synchronization modes. To address the high computational complexity of the proposed method, we introduce HASY, a \qty{6}{\square\mm} application-specific integrated circuit (ASIC) fabricated in \qty{65}{\nm} CMOS technology. HASY generates one \qty{96}{\kHz}, \qty{24}{\bit} alias-free synchronized waveform with up to $512$ harmonics and computes the spectral-resampling transform within only five audio-sample periods.
\end{abstract}

\section{Introduction}
\label{sec:intro}

\begin{figure}[t]
\centering
\includegraphics[width=0.95\columnwidth]{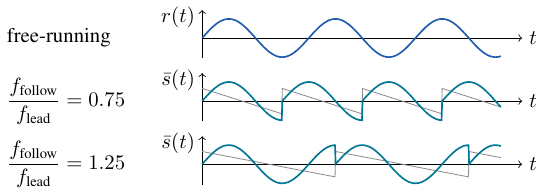}
\caption{\label{fig:hard_sync_illu}{Illustration for hard sync with a sine wave for the following-oscillator waveform. Top: free-running following oscillator. Middle and bottom: (petrol) hard-sync signal and (gray) exemplary leading oscillator.
The middle and bottom each use a different following-to-leading-oscillator frequency ratio.}}
\end{figure}

Oscillator synchronization is a widely used technique in music synthesizers that is able to create rich, expressive waveforms with intricate harmonic content. 
Examples of tracks prominently featuring oscillator synchronization are ``Release the Beast'' by Breakwater~\cite{breakwater1980release} and ``Fletch Theme'' by Harold Faltermeyer~\cite{faltermeyer1985fletch}.
The synthesis method involves two oscillators, a \emph{leading} oscillator and a \emph{following} oscillator, which have fundamental frequencies $\flead$ and~$\ffollow$, respectively.\footnote{In historic terminology, the leading and following oscillators were called ``master'' and ``slave'' oscillators, respectively. }
The classic oscillator-synchronization setup, called hard synchronization (``hard sync'' for short), is illustrated in \fref{fig:hard_sync_illu}.
Whenever the leading oscillator completes one period (i.e., on the rising edge of the waveform), then the following oscillator's phase is reset (i.e., the oscillator starts its waveform anew).
Depending on the following oscillator's amplitude at the reset instant, this reset procedure will introduce a discontinuity in the waveform, causing new high-frequency components that were not present in the free-running following-oscillator waveform. 

\begin{remark}
Conceptually, the leading oscillator's waveform is unimportant.
However, in a traditional analog synthesizer, phase resets take place on rising edges of the leading oscillator.
Thus, we depict the leading oscillator in \fref{fig:hard_sync_illu} with a sawtooth wave.
\end{remark}

Alternative modes to hard sync also exist. Soft oscillator synchronization (``soft sync'' for short) is a generic term for variants that modify the following oscillator's reset procedure in order to control the time-domain discontinuity induced by synchronization.
In practice, soft-sync implementations replace the hard phase reset with rules such as waveform reversal at the reset instant (often referred to as \emph{reversing sync} or \emph{mirrored sync}), or with conditional reset procedures.
These modifications affect the spectral composition of the synchronized waveform and often help reduce the perceived ``harshness'' associated with hard sync. 

\setlength{\textfloatsep}{12pt plus 1pt minus 1pt}
\setlength{\floatsep}{12pt plus 1pt minus 1pt}
\setlength{\dbltextfloatsep}{12pt plus 1pt minus 1pt}
\setlength{\dblfloatsep}{12pt plus 1pt minus 1pt}

\subsection{Digital Implementation of Hard Sync Is Hard}

While analog circuit implementation of oscillator synchronization is rather straightforward, trivial digital emulations, e.g., with a hard phase reset, cause severe aliasing artifacts, as pointed out in~\cite{brandt2001hard}.
The underlying issue is not specific to hard sync: any waveform with discontinuities is not bandlimited, and its high-frequency content aliases when sampled na\"ively.
A brute-force remedy that combines extreme oversampling with a sharp anti-aliasing low-pass filter incurs high computational overhead.
Below, we discuss more efficient approaches that synthesize a bandlimited signal directly; see \cite{valimaki2007antialiasing} for a general overview of anti-aliasing synthesis.

Early work on anti-aliasing synthesis focused on general-purpose techniques that reduce discontinuity-induced aliasing: BLIT~\cite{stilson1996alias} targets classical analog waveforms (sawtooth, square, and triangle); BLEP~\cite{brandt2001hard} was introduced to address hard sync.
A later refinement of BLIT~\cite{nam2009efficient} adds a perceptual analysis of aliasing and applies the method to oscillator synchronization.
More specialized methods have been proposed for particular oscillator types: For example, \cite{lapastina2022general} applies anti-aliasing FIR low-pass filters to a synchronized sine wave, while~\cite{timoney2012virtual} derives an efficient implementation for synchronized sawtooth oscillators from an analytical Fourier-series description. 
However, these approaches rely on waveform-specific derivations, which lack the flexibility of synchronizing \emph{arbitrary} periodic waveforms.
More general anti-aliasing methods, such as polynomial transition regions~\cite{kleimola2011reducing}, have also been used to smooth the discontinuities at each reset.

A more general approach to anti-aliasing is additive synthesis, which enables bandlimited synthesis of arbitrary periodic waveforms. However, an oscillator-based implementation is often considered computationally too expensive for software realizations. Less computationally intensive approaches for bandlimited synthesis instead rely on wavetables (e.g., pre-filtered waveforms at different fundamental frequencies) or methods such as inverse fast Fourier transform (IFFT) synthesis \cite[pp.~458--467]{chamberlin1985musical}, \cite{rodet1992spectral}.
Alternatively, oscillator-based additive synthesis can be implemented in hardware through application-specific integrated circuits (ASICs), as demonstrated in \cite{debernardinis1999efficient,dafx23pmsynth}.
However, as pointed out in \cite{deslauriers2009bandlimited}, a true emulation of oscillator synchronization via additive synthesis would benefit from a spectral transformation describing the synchronized waveform for arbitrary inputs.

\subsection{Contributions}
We propose a general method to digitally emulate alias-free oscillator synchronization for arbitrary periodic free-running following-oscillator waveforms.
In particular, given the Fourier-series coefficients of a bandlimited following-oscillator waveform, we compute the Fourier-series coefficients of the synchronized waveform via a spectral-resampling transform. 
We then generate an alias-free time-domain synchronized signal through additive synthesis.
Our approach naturally supports three different oscillator-synchronization modes: (i) traditional hard sync, (ii) mirrored sync, and (iii) pulsar sync (related to pulsar synthesis~\cite{roads2001pulsar}).
Since the proposed spectral-resampling transform exhibits high complexity, we present HASY (short for hard sync), a real-time digital implementation on an ASIC.
HASY implements spectral resampling and additive synthesis in real time for one following oscillator with up to $512$ harmonics at a sample rate of \qty{96}{\kHz} with \qty{24}{\bit} resolution.

\section{Spectral Resampling and Additive Resynthesis for Oscillator Synchronization}
\label{sec:algorithm}

Our approach for alias-free oscillator synchronization operates in the Fourier-series domain.
Given a finite set of Fourier-series coefficients $\{\acfollow{c}_n\}_{n=-N}^{N}$ describing a bandlimited free-running periodic following-oscillator waveform, we compute the corresponding Fourier-series coefficients $\{\achard{c}_n\}_{n=-N}^{N}$ of the synchronized waveform with leading-oscillator period $\Tlead=1/\flead$.
We call this linear mapping \emph{spectral resampling}, which forms the core idea of our approach.
With the transformed Fourier-series coefficients, we then generate the synchronized time-domain signal via additive synthesis\,---\,here, we restrict the number of synthesized harmonics as a function of $\Tlead$ and the sample rate $\fs$ to avoid aliasing.

\begin{remark}
Our approach computes the periodic waveform resulting from oscillator synchronization and does \emph{not} implement a phase-synchronous reset in the time domain.
\end{remark}

\begin{remark}
If $\,\Tlead=1/\flead$ is modulated rapidly over time, then our approach may still induce aliasing artifacts each time the transform is recomputed. 
\end{remark}

We now detail our method for hard sync and then outline the other two synchronization modes: mirrored sync and pulsar sync.

\subsection{Spectral Resampling for Hard Sync}
\label{sec:hard_sync}

Let $\{\acfollow c_n\}_{n=-N}^{N}$ be a set of complex-valued Fourier-series coefficients that describe the output signal from a free-running (i.e., unsynchronized) following oscillator with period $\Tfollow$:
\begin{align} \label{eq:complexFourierseries}
r(t) = \sum_{n=-N}^N \acfollow c_n \erm^{\jrm\frac{2\pi}{\Tfollow}nt}, \quad t \in \mathbb{R}.
\end{align}

\begin{remark}
\label{rm:bandlimited-following-osc}
Since $N$ in \fref{eq:complexFourierseries} is finite, the free-running following-oscillator signal $r(t), t \in \mathbb{R}$ is bandlimited to $N$ harmonics. This also implies that our approach will emulate oscillator synchronization applied to an already bandlimited following-oscillator waveform.
Note that we use $r(\cdot)$ to denote $r(t), t \in \mathbb{R}$.
\end{remark}

\begin{remark}
We rely on the complex-coefficient formulation to simplify our derivations. Since the produced time-domain signals $r(\cdot)$ are assumed to be real-valued, their Fourier-series coefficients $\{\acfollow c_n\}_{n=-N}^{N}$ exhibit complex-conjugate symmetry $c_{-n}=c_n^*$. At the end of our calculations, we convert the complex-valued Fourier-series coefficients to the real-valued Fourier-series coefficients $\big\{a_0,\{a_n, b_n\}_{n=1}^{N}\big\}$ through the relationships $a_0=c_0$, $a_n=2\Re\{c_n\}$ and $b_n=-2\Im\{c_n\}$, where $\Re\{c_n\}$ and $\Im\{c_n\}$ are the real and imaginary parts of $c_n$, respectively.
\end{remark}

We now aim to emulate hard sync with a leading-oscillator period $\Tlead$.
In the Fourier-series coefficient domain, only the ratio of $\ffollow$ to $\flead$ matters. Therefore, we introduce the period ratio 
\begin{align}
P \define \frac{\Tlead}{\Tfollow} = \frac{\ffollow}{\flead}.
\end{align}
Without loss of generality, we can normalize the periods to
\begin{align}
  \Tfollow=1 \quad \text{and} \quad \Tlead=P, \label{eq:period_normalization}
\end{align}
which will simplify our derivations. 
With this normalization, we divide the Fourier-series coefficient transform into two parts: (i) a pre-rotation followed by (ii) the spectral-resampling transform.

\subsubsection{Pre-Rotation}
\label{sec:prerotation}

\begin{figure}[tp]
\centerline{\includegraphics[width=0.95\columnwidth]{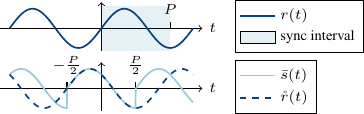}}
\caption{\label{fig:time_shift}{Illustration of the time shift operation with a sine wave as following-oscillator waveform and $P=0.75$. Top: free-running following-oscillator waveform $r(t)$ with the desired sync interval marked. Bottom: time-shifted following-oscillator signal~$\acrot r(t)$ and resulting hard-sync signal~$\achard s(t)$ for alignment with the symmetrical Fourier integration window.}}
\end{figure}

When illustrating oscillator synchronization, one usually considers the leading-oscillator period $[0, \Tlead]$, which becomes $[0,P]$ after the normalization in \fref{eq:period_normalization}.
However, the spectral-resampling calculations simplify if we consider the interval $[-P/2,P/2]$ as the leading-oscillator period.
To emulate oscillator synchronization correctly, the start of the following-oscillator waveform should align with the start of the leading-oscillator period (see \fref{fig:time_shift}, top). 
To this end, we time-shift the following-oscillator signal as $\acrot{r}(t) \define r(t - \tau)$ with $\tau = -P/2$ (see \fref{fig:time_shift}, bottom).
In the Fourier-series coefficient domain, this time shift corresponds to a phase shift, and the ``pre-rotated'' coefficients $\{\acrot c_n\}_{n=-N}^N$ associated with the time-shifted signal $\acrot{r}(\cdot)$ are given by
\begin{align} \label{eq:prerotation}
\acrot c_n = \acfollow c_n \erm^{-\jrm 2\pi \tau n/\Tfollow} = \acfollow c_n \erm^{-\jrm 2\pi \tau n}, \quad n = -N, \ldots, N. 
\end{align}
Note that we can apply the phase shift from \fref{eq:prerotation} to the real and imaginary parts of the Fourier-series coefficients as 
\begin{align}
\left[\begin{array}{c}
\!\!\Re\{\acrot{c}_n\} \!\! \\
\!\!\Im\{\acrot{c}_n\} \!\!
\end{array}\right]  &=
\left[\begin{array}{cc}
\!\!\phantom{-}\cos(2\pi n \tau) & \sin(2\pi n \tau) \!\!\\
\!\!-\sin(2\pi n \tau) & \cos(2\pi n \tau)\!\!
\end{array}\right]
\left[\begin{array}{c}
\!\!\Re\{c_n\} \!\!\\
\!\!\Im\{c_n\}\!\!
\end{array}\right]\!.\label{eq:prerotate_real}
\end{align}

\subsubsection{Spectral Resampling}
\label{sec:spectral_resampling_hard}

We are now ready to calculate the complex-valued Fourier-series coefficients $\{\achard{c}_n\}_{n=-N}^{N}$ of the same signal $\acrot{r}(\cdot)$ but over a new period $\Tlead=P$ instead of $\Tfollow=1$.
This operation is equivalent to synchronizing (resetting) the following oscillator at a rate of $1/P$, which will produce the desired hard-sync oscillator signal~$\achard{s}(\cdot)$. 
The $n$th complex-valued Fourier-series coefficient associated with the synchronized waveform $\achard{s}(\cdot)$ is calculated as
\begin{align}
\achard{c}_n & = \frac{1}{P} \int_{-P/2}^{P/2} \acrot r(t)\erm^{-\jrm\frac{2\pi}{P}nt} \text{d}t \\
& = \frac{1}{P} \int_{-P/2}^{P/2}  \sum_{k=-N}^N \acrot c_k \erm^{\jrm{2\pi}kt} \erm^{-\jrm\frac{2\pi}{P}nt} \text{d}t \\
& =  \sum_{k=-N}^N \acrot c_k \frac{1}{P} \int_{-P/2}^{P/2}  \erm^{\jrm2\pi( k - \frac{n}{P})t} \text{d}t \\
& =  \sum_{k=-N}^N \acrot c_k  \sinc\!\left(n-kP\right)\!,  \label{eq:transform_hard}
\end{align}
where we define the sinus cardinalis (sinc) function as
\begin{align} \label{eq:sinc}
\sinc(x) \define  \left\{\begin{array}{ll}
\frac{\sin(\pi x)}{\pi x} & \text{if  } x\neq0 \\
1 & \text{if  } x=0.
\end{array}\right.
\end{align}

In essence, the spectral-resampling transform in~\fref{eq:transform_hard} takes in all $2N+1$ Fourier-series coefficients $\{\acrot{c}_k\}_{k=-N}^N$ and produces a new set of Fourier-series coefficients $\{\achard{c}_n\}_{n=-N}^N$, which describe the hard-sync waveform with the leading-to-following-oscillator period ratio~$P$.
Mathematically, \fref{eq:transform_hard} corresponds to a \emph{linear} transform that maps the Fourier-series coefficients of the pre-rotated (i.e., time-shifted) following-oscillator waveform $\{\acrot c_n\}_{n=-N}^N$ to the coefficients of the hard-synchronized waveform $\{\achard{c}_n\}_{n=-N}^N$.

\begin{remark}
In \fref{eq:transform_hard}, we map $2N+1$ input to $2N+1$ output coefficients. Alternatively, one could compute an output coefficient set of different size by evaluating \fref{eq:transform_hard} for $n=-N',\ldots,N'$.
This would enable individual control of the input and output bandlimiting.
\end{remark}


In audio applications, one is typically interested in synthesizing a real-valued time-domain signal.
To this end, we convert the complex-valued Fourier-series coefficients $\{\achard{c}_n\}_{n=-N}^{N}$ into the corresponding real-valued Fourier-series coefficients $\big\{\achard{a}_0,\{\achard{a}_n,\achard{b}_n\}_{n=1}^N\big\}$, removing the complex conjugate redundancy of real-valued signals.
Instead of first computing the complex-valued coefficients and then transforming them into the real domain, we can directly compute the linear spectral-resampling transform on the real-valued coefficients for $n=1,\ldots,N$:
\begin{align}
\achard{a}_0 & = \acrot{a}_0 + \sum_{k=1}^N \acrot{a}_k \sinc(kP) \label{eq:transform_hard_a0} \\
\achard{a}_n & = \sum_{k=1}^N \acrot{a}_k \bigl(\sinc(n-kP)+\sinc(n+kP)\bigr) \label{eq:transform_hard_a} \\
\achard{b}_n & = \sum_{k=1}^N \acrot{b}_k \bigl(\sinc(n-kP)-\sinc(n+kP)\bigr).\label{eq:transform_hard_b}
\end{align}
Note that the pre-rotation (time shift) must first be carried out using the explicit method shown in \fref{eq:prerotate_real}, which can also be applied to the real-valued Fourier-series coefficients.

\subsection{Additive Resynthesis}
\label{sec:resynthesis}


With the real-valued Fourier-series coefficients of the hard-sync output signal, we synthesize an alias-free time-domain signal using the approach from \cite{dafx23pmsynth}.
Concretely, the alias-free discrete-time hard-sync output signal $\achard{s}[\cdot]$ at sample rate $\fs$ is given by
\begin{align}\label{eq:add_synth}
\achard s\left[\ell\right] = \achard a_0 + \!\!\sum_{n \in \setN}
\achard{a}_n \cos\!\left(\frac{2\pi n}{\Tlead \fs}\ell\right)\!
+ \achard{b}_n \sin\!\left(\frac{2\pi n}{\Tlead \fs}\ell\right)\!,
\end{align}
where we only consider harmonics in the set
\begin{align}
\setN \define \{n = 1,\ldots,N : n/\Tlead < \fs/2\}. \label{eq:nyquist}
\end{align}
In other words, we only sum harmonics below the Nyquist frequency $\fs/2$.
This is why our approach is alias-free.

\begin{remark}
In \fref{eq:add_synth}, we reintroduced the period $\Tlead$ of the leading oscillator, abandoning the normalization in \fref{eq:period_normalization} and setting $\Tlead$ to arrive at the desired leading-oscillator fundamental frequency.
\end{remark}

\begin{remark}
In contrast to the additive-synthesis approach in~\cite{dafx23pmsynth}, we do not allow arbitrary partials. Instead, we restrict ourselves to pure harmonics, i.e., integer multiples of the fundamental.
\end{remark}

\begin{remark}
Due to the pre-rotation (time shift) from \fref{sec:prerotation}, the reset instant in the hard-sync output signal takes place at $t = -\Tlead/2$, i.e., the output signal appears time-shifted by $\tau = -P/2$.
If such a time shift is undesired, one could either apply a post-rotation to the coefficients $\{\achard{c}_n\}$ or undo the time shift during additive synthesis \fref{eq:add_synth}. In our implementation, we ignore such a post-processing step  entirely.
\end{remark}

\subsection{Spectral Resampling for Mirrored Sync}
\label{sec:mirrored_sync}

\begin{figure}[t]
\centerline{\includegraphics[width=0.95\columnwidth]{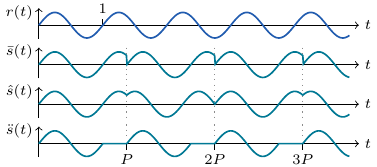}}
\caption{\label{fig:soft_sync}{Comparison of different sync modes, here using a sinusoidal following-oscillator waveform and $P = \frac{11}{8}$. From top to bottom: (free-running) following oscillator $r(t)$, hard sync $\achard s(t)$, mirrored sync $\acmirror s(t)$, and pulsar sync $\acpulsar s(t)$.}}
\end{figure}

Our resampling method naturally extends to other oscillator-synchronization methods, such as \emph{mirrored sync}. Instead of resetting the following oscillator as for hard sync, mirrored sync ``reflects'' the following-oscillator waveform at the reset instant, as illustrated in \fref{fig:soft_sync}.
Mathematically, the mirrored-sync output signal $\acmirror{s}(\cdot)$ has a total period of $2\,\Tlead$ ($\Tlead$ for the forward and $\Tlead$ for the mirrored waveform) and can be described as
\begin{align}
\acmirror{s}(t) & = \left\{\begin{array}{ll}
r(t)                & \text{if }\,\, 0 \leq t < \Tlead \\
r(2\Tlead-t)  & \text{if }\,\, \Tlead \leq t < 2\,\Tlead.
\end{array}\right. \label{eq:mirrored_sync_time}
\end{align}

Following the same approach as in \fref{sec:hard_sync} for hard sync, we can derive the complex-valued Fourier-series coefficient transform for the mirrored sync.
Again, we assume $\Tfollow=1$ and $\Tlead=P$.
Starting from the real-valued coefficients $\big\{{a}_0,\{{a}_n,{b}_n\}_{n=1}^{N}\big\}$, we first apply the pre-rotation in~\fref{eq:prerotate_real} with $\tau=-P$ to get $\big\{\acrot{a}_0,\{\acrot{a}_n,\acrot{b}_n\}_{n=1}^{N}\big\}$.
We then derive the spectral-resampling transform for the mirrored sync analogous to  \fref{sec:spectral_resampling_hard}, yielding the coefficients $\big\{\acmirror{a}_0,\{\acmirror{a}_n,\acmirror{b}_n\}_{n=1}^{N}\big\}$ that describe the mirrored-sync signal~$\acmirror{s}(\cdot)$.
The resulting transform for real-valued Fourier-series coefficients is 
\begin{align}
&\begin{aligned}
\acmirror{a}_0  &= \acrot{a}_0 + \sum_{k=1}^N \acrot{a}_k \sinc(2kP) - \acrot{b}_k \versinc(2kP) \label{eq:transform_mirrored_a0}
\end{aligned}\\
&\begin{aligned}
 \acmirror{a}_n\!=\!\sum_{k=1}^N \biggl[ & \acrot{a}_k \bigl(\sinc(n-2kP) +\sinc(n+2kP)\bigr) + \\
 & \acrot{b}_k \bigl(\versinc(n-2kP)-\versinc(n+2kP)\bigr) \biggr] \label{eq:transform_mirrored_a}
\end{aligned}\\
&\begin{aligned}
\acmirror{b}_n  &= 0, \label{eq:transform_mirrored_b}
\end{aligned}
\end{align}
for $n=1,\ldots,N$. Here, we define the \emph{versed sinus cardinalis} (versinc) function\footnote{We chose this terminology in analogy to the sinc function by using the  \emph{versed sine} function $\mathrm{versin}(\phi)=1-\cos(\phi)$ instead.} as follows:
\begin{align} \label{eq:versinc}
\versinc(x) \define  \left\{\begin{array}{ll}
\frac{1 - \cos(\pi x)}{\pi x} & \text{if } x\neq0 \\
0 & \text{if } x=0.
\end{array}\right.
\end{align}

\begin{remark}
In contrast to \fref{eq:transform_hard_a0}\,--\,\fref{eq:transform_hard_b} for hard sync, the resulting linear transform in \fref{eq:transform_mirrored_a0}, \fref{eq:transform_mirrored_a} also features $\versinc(\cdot)$ terms. 
Nonetheless, the implementation similarity between the sinc and versinc functions enables, as discussed in \fref{sec:implementation}, a shared hardware implementation of both hard and mirrored sync.
\end{remark}

\begin{remark}
The Fourier-series coefficients $\{\acmirror{b}_n\}_{n=1}^{N}$ in  \fref{eq:transform_mirrored_b} vanish, which is a direct consequence of the even symmetry in the mirrored-sync signal $\acmirror{s}(\cdot)$.
\end{remark}

In order to arrive at an alias-free output signal $\acmirror{s}[\ell]$ for the mirrored sync, we use the same additive-synthesis approach from \fref{sec:resynthesis} with the Fourier-series coefficients $\big\{\acmirror{a}_0,\{\acmirror{a}_n,\acmirror{b}_n\}_{n=1}^{N}\big\}$. However, since $\acmirror{s}(\cdot)$ has a period of $2\,\Tlead$, one must utilize $2\,\Tlead$ instead of $\Tlead$ in \fref{sec:resynthesis}. 

\subsection{Spectral Resampling for Pulsar Sync}
\label{sec:pulsar_sync}

Another alternative that fits naturally in our spectral-resampling framework is what we call \emph{pulsar sync}.
Instead of resetting the following oscillator as hard sync does, pulsar sync periodically mutes the following oscillator after each following-oscillator period \Tfollow.
Therefore, this mode only makes sense for $P\geq1$.
Mathematically, one period of the pulsar-sync output signal $\acpulsar{s}(t)$ with period $\Tlead$ can be described as
\begin{align}
\acpulsar{s}(t) & = \left\{\begin{array}{ll}
r(t)  & \text{if } \,\, 0 \leq t < \Tfollow \\
0     & \text{if } \,\, \Tfollow \leq t < \Tlead.
\end{array}\right.
\end{align}
Figure~\ref{fig:soft_sync} features an illustration of the resulting waveform.
Note that pulsar sync is conceptually equivalent to \emph{pulsar synthesis}~\cite{roads2001pulsar} (hence the name), with the following-oscillator waveform serving as the \emph{pulsaret} in combination with a rectangular \emph{pulsaret window}.

By following the same approach as in \fref{sec:hard_sync} for hard sync, we can derive the complex-valued Fourier-series coefficient transform for the pulsar sync.
Again, we assume \mbox{$\Tfollow=1$} and \mbox{$\Tlead=P$}. Since we are considering audio signals, we will assume that the following-oscillator waveform has no DC component, so that $c_0\!=\!a_0\!=\!\acrot{a}_0\!=\!0$. This will simplify some of the resulting expressions.
Starting from the real-valued coefficients $\big\{{a}_0,\{{a}_n,{b}_n\}_{n=1}^{N}\big\}$,  we first apply the pre-rotation in \fref{eq:prerotate_real} with \mbox{$\tau=-1/2$} to get $\big\{\acrot{a}_0,\{\acrot{a}_n,\acrot{b}_n\}_{n=1}^{N}\big\}$.
We then derive the spectral-resampling transform for the pulsar sync analogous to the procedure in \fref{sec:spectral_resampling_hard}, yielding the coefficients $\big\{\acpulsar{a}_0,\{\acpulsar{a}_n,\acpulsar{b}_n\}_{n=1}^{N}\big\}$ that describe the pulsar-sync signal $\acpulsar{s}(\cdot)$.
The resulting transform for real-valued Fourier-series coefficients is 
\begin{align}
\acpulsar{a}_0 &= \frac{\acrot{a}_0}{P} = 0 \label{eq:transform_pulsar_a0} \\
\acpulsar{a}_n &= \sum_{k=1}^N \frac{\acrot{a}_k}{P}\!\left(\sinc\left(\frac{n}{P}-k\right)+\sinc\left(\frac{n}{P}+k\right)\right) \label{eq:transform_pulsar_a}  \\
\acpulsar{b}_n &= \sum_{k=1}^N \frac{\acrot{b}_k}{P}\!\left(\sinc\left(\frac{n}{P}-k\right)-\sinc\left(\frac{n}{P}+k\right)\right), \label{eq:transform_pulsar_b}
\end{align}
for $n=1,\ldots,N$. 

In order to arrive at an alias-free output signal $\acpulsar{s}[\ell]$ for the pulsar sync, we use the same additive-synthesis approach from \fref{sec:resynthesis} with the Fourier-series coefficients $\big\{\acpulsar{a}_0,\{\acpulsar{a}_n,\acpulsar{b}_n\}_{n=1}^{N}\big\}$.

\begin{remark}
Although derived for $P\geq1$, the spectral transform \fref{eq:transform_pulsar_a}, \fref{eq:transform_pulsar_b} appears to extend to $P<1$, where neighboring following-oscillator-waveform cycles overlap, which is similar to \emph{overlapped pulsaret-width modulation}~\cite[p.~136]{roads2001pulsar}. A more detailed analysis of this case is left for future work.
\end{remark}

\section{ASIC Implementation}
\label{sec:implementation}

While our described spectral-resampling approach provides an alias-free formulation of oscillator synchronization, it comes with two practical drawbacks: (i)~the coefficient transform has a computational complexity of $\mathcal{O}(N^2)$ and (ii)~computing the transform coefficients requires evaluating many terms that consist of trigonometric functions and divisions (e.g., sinc or versinc functions).
For example, the hard-sync spectral transform \fref{eq:transform_hard_a}, \fref{eq:transform_hard_b} with $N=512$ requires $512^2 \cdot 2 = 524\,288$ sinc function evaluations, each involving a sine function evaluation and a division.
For real-time sound synthesis in software with modulated period ratio\footnote{Modulation of $\flead$ or $\ffollow$ implies a modulation of $P$.}, even a small number of harmonics quickly results in prohibitive complexity.
As a reference, an unoptimized NumPy implementation requires between \qty{2.7}{\ms} and \qty{5.8}{\ms} to compute the spectral-resampling transform for a sawtooth following-oscillator waveform and $N=512$ (depending on the sync mode and the period ratio)\footnote{Measured on a MacBook Pro (M2 Pro). Generally, the mirrored-sync resampling transform consistently took longer than the other two sync modes, likely due to the custom versinc function evaluation.}, which corresponds to a hypothetical maximum update rate between \qty{172}{\Hz} and \qty{370}{\Hz}.
At these update rates, high modulation rates or depths may introduce audible artifacts each time the Fourier-series coefficients are updated, so a faster computation of the transform is desirable.
To this end, we harness the efficiency of ASIC implementations that can be specialized to the specific computations required by our approach.
Concretely, we propose an ASIC called \emph{HASY}, which implements the three proposed spectral-resampling methods (hard, mirrored, and pulsar sync) and additive synthesis to enable real-time alias-free oscillator synchronization with $N=512$ harmonics.

\begin{figure}[tp]
\centering
\includegraphics[scale=1.0 ]{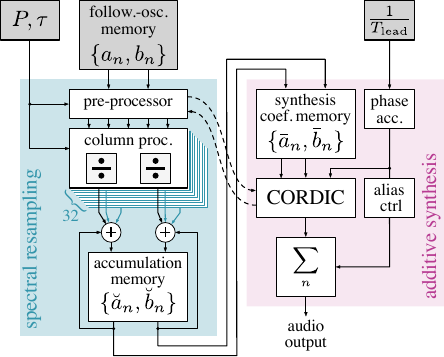}
\caption{\label{fig:hasy_overview}Architecture overview of the HASY ASIC with spectral-resampling engine and additive-synthesis oscillator. Gray elements are user-configurable.}
\end{figure}

\subsection{Architecture Overview}

HASY is implemented as custom digital hardware using a fixed-point datapath (bit-widths of order \qty{24}{\bit}) and a parallel architecture to enable real-time execution of the proposed spectral-resampling approach.
\fref{fig:hasy_overview} provides the architecture overview of HASY:
The left side of the block diagram depicts the \emph{spectral-resampling engine} that implements the Fourier-coefficient transforms for hard sync, mirrored sync, and pulsar sync described in Sections~\ref{sec:hard_sync}, \ref{sec:mirrored_sync}, and \ref{sec:pulsar_sync}, respectively.
The right side depicts the \emph{additive-synthesis oscillator} that implements the alias-free synthesis of \fref{sec:resynthesis} and is realized as a stripped-down version of the big Fourier oscillator (BFO) \cite{dafx23pmsynth}.

\subsubsection{Sound Parameter Inputs}

The intended application scenario for HASY is a digital hardware synthesizer system with a central host microcontroller.
In this scenario, the host controls HASY via an SPI interface (not shown in \fref{fig:hasy_overview}), which allows the host to set the parameters that determine the sound produced by HASY.
Specifically, the SPI interface allows the host to set the following-oscillator Fourier-series coefficients $\{a_n, b_n\}_{n=1}^{512}$ (HASY assumes $a_0\!=\!0$, which is often reasonable for audio signals as they have no DC component), the leading-to-following-oscillator period ratio~$P$, and the leading-oscillator fundamental frequency $\flead$, each with \qty{24}{\bit} resolution.

\subsubsection{Spectral-Resampling Engine}

The spectral-resampling engine consists of four main parts (left side in \fref{fig:hasy_overview}, top to bottom): (i)~following-oscillator Fourier-series coefficient memory (short \emph{following-oscillator memory}), (ii)~pre-processor, (iii)~parallel column processors, and (iv)~accumulation Fourier-series coefficient memory (short \emph{accumulation memory}). Those blocks are now discussed in more detail:

(i) The following-oscillator memory contains the real-valued Fourier-series coefficients $\{a_n, b_n\}_{n=1}^{512}$ that are received via SPI.

(ii) When the spectral-resampling engine is running, the \emph{pre-processor} sequentially fetches coefficient pairs $\{a_n, b_n\}$ from the following-oscillator memory and applies the pre-rotation in~\fref{eq:prerotate_real}.
Besides this, the pre-processor also evaluates trigonometric functions that appear in the spectral-resampling transforms (e.g., $\sin(\pi(n \pm kP))$ for hard sync).
These trigonometric functions are computed using a \emph{coordinate rotation digital computer (CORDIC)}\footnote{CORDICs allow for efficient yet accurate computation of trigonometric functions.}~\cite{volder1959cordic} that is time-shared with the additive-synthesis side.
To further optimize the computation of the trigonometric terms, symmetries of the trigonometric functions are exploited.
Ultimately, the pre-processor passes the rotated coefficient pairs $\{\acrot a_n, \acrot b_n\}$ along with the computed trigonometric values to the parallel column processors.

(iii) An array of $32$~parallel \emph{column processors} is used to compute the terms of the $\sum$-sums found in the spectral-resampling transforms.
These terms from the spectral-resampling transforms contain quotients (e.g., the fraction that appears in $\sinc(n \pm kP)$ for hard sync).
In order to compute these quotients, custom divider units are implemented using the so-called \emph{division-by-hand} algorithm.
To avoid numerical issues with very small denominators, the dividers use a zeroth-order or first-order Taylor expansion\footnote{The hard-sync transform \fref{eq:transform_hard_a}, \fref{eq:transform_hard_b} and pulsar-sync transform \fref{eq:transform_pulsar_a}, \fref{eq:transform_pulsar_b} feature the $\sinc(\cdot)$ quotient, where we use a zeroth-order Taylor approximation. The mirrored-sync transform \fref{eq:transform_mirrored_a} features $\versinc(\cdot)$, where we use a first-order Taylor approximation.} for denominators smaller than $2^{-12}$.
The column processors also implement the summation of the quotients (resp. subtraction) that appear in the spectral-resampling transforms.
After this, the column processors carry out the multiplication with the input Fourier-series coefficients.
Then, the column processors pass the terms of the $\sum$-sums found in the spectral-resampling transforms to the accumulation memory.

(iv) Finally, the \emph{accumulation memory} is dedicated to the accumulation of the terms from the $\sum$-sums in the linear spectral-resampling transform.

\subsubsection{Real-Time Capability}

HASY is designed to run at a clock frequency $f_\text{clk}$ that is a fixed multiple of the audio output sample rate $\fs$, i.e., $f_\text{clk} = 2048\,\fs$.
Thus, for the target audio sample rate of \qty{96}{\kHz}, HASY requires a target clock frequency of $f_\text{clk} \approx \qty{196.6}{\MHz}$.
Thanks to the parallel architecture of the spectral-resampling engine, which comprises $32$~parallel column processors, the Fourier-series coefficient transform for $N=512$ harmonics is completed in only $8\,240$ clock cycles, independently of the sync mode.
Thus, at the target clock frequency, the spectral-resampling transform takes approximately \qty{42}{\us}, which corresponds to just over $4$ audio-sample periods at \qty{96}{\kHz}. 
Thus, if desired, HASY can execute a new spectral-resampling transform every $5$ audio samples, which corresponds to a parameter update rate of approximately \qty{19.2}{\kHz}.
Such a high parameter update rate makes HASY real-time capable, and modulating the leading oscillator frequency $\flead$ is possible. 

\subsubsection{Alias-Free Additive-Synthesis Oscillator}

As shown on the right side in \fref{fig:hasy_overview}, HASY also contains a stripped-down version of a BFO \cite{dafx23pmsynth} that synthesizes one voice of the alias-free time-domain audio output signal.

At the end of the spectral-resampling computation, the resulting Fourier-series coefficients are written directly to the \textit{synthesis coefficient memory}.
The desired leading-oscillator pitch $1/\Tlead$ is set by the system host via SPI.
The alias-free additive synthesis in \fref{eq:add_synth} is implemented with a CORDIC that computes the $\sin(\cdot) + \cos(\cdot)$ terms in combination with a conditional accumulation to ensure the Nyquist condition \fref{eq:nyquist} is met.
Note that in contrast to \fref{eq:add_synth}, the DC component $\achard{a}_0$ is ignored, since it is considered irrelevant for the intended application.
The additive synthesizer uses a pipelined architecture with a throughput of one coefficient pair per clock cycle.
With $N=512$ harmonics, the additive synthesis occupies only about one fourth of the available $2048$ clock cycles per audio sample.
Thus, when the synthesis coefficient memory is receiving newly computed Fourier-series coefficients from the spectral-resampling engine, the additive-synthesis computation is scheduled accordingly to avoid memory-access contention.
Finally, the output time-domain signal is passed to an I2S interface with \qty{96}{\kHz} and \qty{24}{\bit} resolution.

In contrast to the BFO ASIC~\cite{dafx23pmsynth}, HASY contains only one (mono) oscillator and the number of harmonics is reduced from $1024$ to $512$.
Also, the HASY design operates at a reduced resolution in the datapath from \qty{32}{\bit} to \qty{24}{\bit} to reduce silicon area. 

\subsection{ASIC Measurement Results and Comparison}

\begin{figure}[tp]
\centerline{\includegraphics[width=0.45\columnwidth]{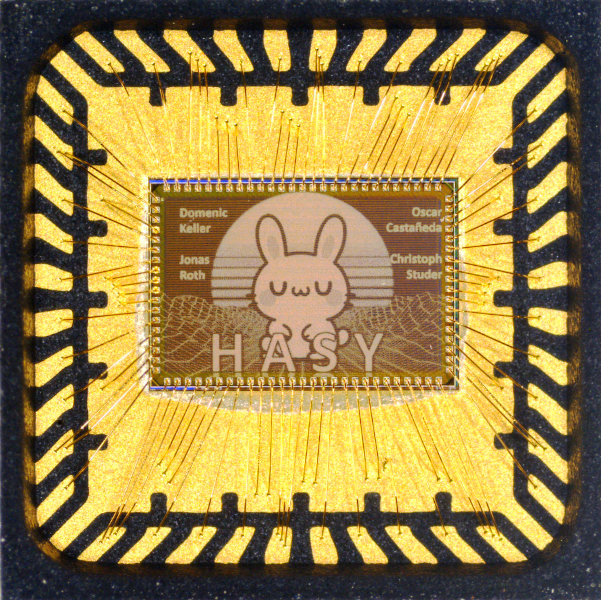}}
\caption{\label{fig:asic_die}{ASIC micrograph. The photo shows the HASY chip inside its package (lid removed). The logo is implemented through the so-called metal fill in the top layer of the chip.}}
\end{figure}

\fref{fig:asic_die} shows a micrograph of the \qty{6}{\square\mm} HASY ASIC, which was fabricated in TSMC 65\,nm LP CMOS technology.\footnote{While ASIC design is known to be costly, HASY was fabricated through a multi-project wafer service, where the expensive mask costs are shared between multiple designs. For HASY, 100 chips were fabricated and 10 packaged for a total cost of $\text{\texteuro}\,24\,000$, a process that took five months. The design of the HASY ASIC was enabled by the experience of our research group and the Microelectronics Design Center at ETH Zurich (see the IIS Chip Gallery: \url{http://asic.ethz.ch}).}
The ASIC utilizes a QFN40 package and was successfully tested for the hard-sync mode with a sinusoidal following-oscillator waveform with $P=0.5$ and $\flead=\qty{3}{\kHz}$.
This scenario was also used for the following measurements: At the nominal \qty{1.2}{\V} core supply and \qty{300}{K} room temperature, the ASIC achieves a maximum measured clock frequency of \qty{250}{\MHz}, exceeding the required \qty{196.6}{\MHz} to support an audio sample rate of $\fs = \qty{96}{\kHz}$.
The measured power consumption at \qty{200}{\MHz} clock frequency is \qty{242}{\mW}.

\fref{tbl:specs} summarizes the key implementation characteristics of HASY alongside those of the BFO additive oscillator in~\cite{dafx23pmsynth}.
Clearly, the support of oscillator synchronization comes at a premium. The BFO chip generates a total of $1024\cdot 4 \cdot 2=8192$ partials per sample in only \qty{3}{\square\mm}, but does not support oscillator synchronization. In contrast, HASY generates $512$ harmonics per sample in \qty{6}{\square\mm}, while supporting three distinct oscillator-synchronization modes.  
This area overhead is almost entirely attributable to the spectral-resampling engine, which accounts for $97\%$ of HASY's cell area, while the additive-synthesis oscillator occupies less than $3\%$.

We note that the current implementation of HASY contains errors that escaped verification and limit the functionality of the fabricated HASY ASIC.
These errors are located in the control logic and affect part of the configuration interface.
Nevertheless, the fixed-point golden model together with the ASIC measurements carried out in functioning modes of the chip confirm the core architectural concept and demonstrate that the ASIC achieves real-time processing capability (cf. \fref{tbl:specs}).
A fully functional reimplementation of the ASIC is ongoing work.

{\captionsetup[subfigure]{skip=4pt}
\begin{figure*}[t]
\centering
\begin{subfigure}[t]{0.49\textwidth}
  \centering
  \includegraphics[width=\linewidth]{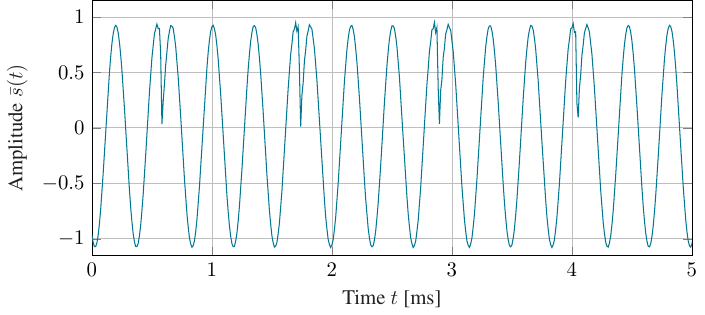}
  \caption{sine hard sync: time domain}
\end{subfigure}\hfill
\begin{subfigure}[t]{0.49\textwidth}
  \centering
  \includegraphics[width=\linewidth]{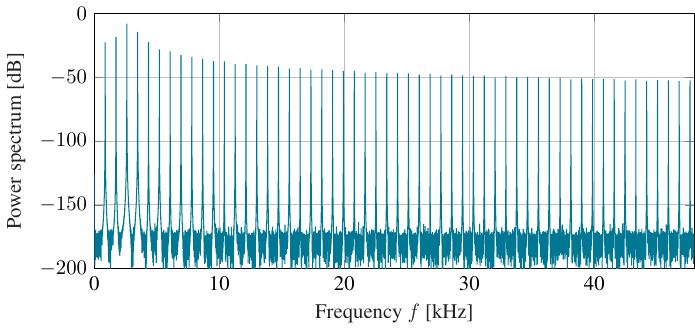}
  \caption{sine hard sync: spectrum}
\end{subfigure}
\begin{subfigure}[t]{0.49\textwidth}
  \centering
  \includegraphics[width=\linewidth]{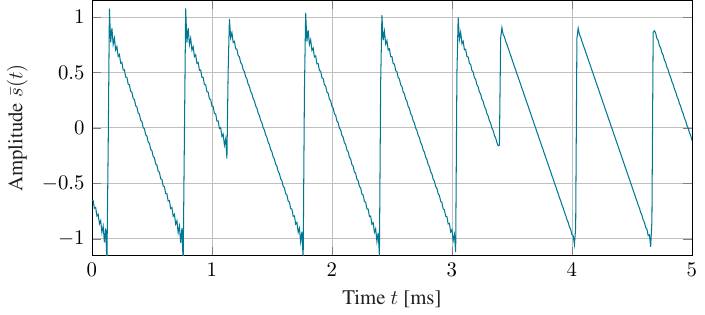}
  \caption{sawtooth hard sync: time domain}
\end{subfigure}\hfill
\begin{subfigure}[t]{0.49\textwidth}
  \centering
  \includegraphics[width=\linewidth]{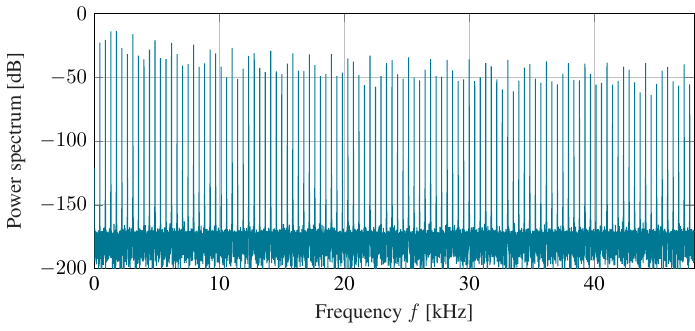}
  \caption{sawtooth hard sync: spectrum}
\end{subfigure}
\caption{\label{fig:eval_waveform_grid}Hard-sync output generated by the HASY fixed-point golden model in time- (left) and frequency- (right) domains. 
The sine hard-sync case (top row) uses $\ffollow = 2\,900.33$\,Hz and $\flead = 866.42$\,Hz, as reported in \cite[Figure 2]{lapastina2022general}.
The sawtooth hard-sync case (bottom row) uses $\ffollow = 1\,575$\,Hz and $\flead = 441$\,Hz, as reported in \cite[Figure 6]{timoney2012virtual}.
}
\end{figure*}
}
    
\begin{figure}[t]
\centering
\begin{subfigure}[b]{0.95\columnwidth}
\includegraphics[width=\textwidth]{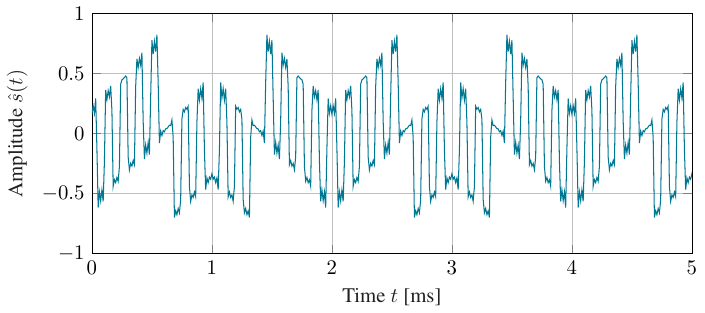}
\caption{Mirrored-sync sandstorm waveform.}
\end{subfigure}
\begin{subfigure}[b]{0.95\columnwidth}
\includegraphics[width=\textwidth]{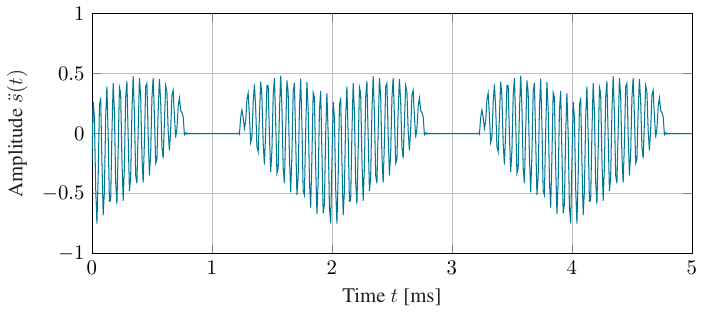}
\caption{Pulsar-sync heart waveform.}
\end{subfigure}
\caption{\label{fig:sand_heart}{Soft-sync variants applied to unconventional waveforms. The sandstorm waveform is inspired by the one used in Darude's ``Sandstorm''~\cite{darude1999sandstorm}.}}
\end{figure}


\begin{table}[t]
\centering
  \caption{\label{tbl:specs}Comparison of HASY and BFO ASICs.}
	\begin{minipage}{0.99\columnwidth}
    \renewcommand{\footnoterule}{}
    \centering
	\begin{tabular}{@{}lcc@{}}
		\toprule
                                & \multicolumn{1}{c}{this work}   & \multicolumn{1}{c}{\cite{dafx23pmsynth}} \\
                                & \multicolumn{1}{c}{HASY}        & \multicolumn{1}{c}{BFO}\\
    \midrule
    Oscillator synchronization  &  yes                             & no \\    
    Voices                      & $1$                             & $2$ \\
    Oscillators per voice       & $1$\footnote{Our approach to hard sync requires only one oscillator.}                             & $4$ \\
    Harmonics per oscillator    & $512$                           & $1024$ \\

    Max.\ clock frequency        & \qty{250}{\MHz}                 & \qty{154}{\MHz} \\
    Chip area                   & \qty{6}{\square\mm}             & \qty{3}{\square\mm}\\
    Power consumption           & \qty{242}{\mW}                  & \qty{178}{\mW}\\
    \bottomrule
	\end{tabular}
    \end{minipage}
\end{table}

\section{Evaluation}

We now evaluate the accuracy of the proposed hard-sync synthesis method and its hardware implementation on the HASY ASIC. For brevity, we restrict our evaluation to the hard-sync mode.

\subsection{Baselines and Performance Metric}

To assess the accuracy of the proposed ASIC, we use our fixed-point golden model implemented in MATLAB. This model is bit-true to the HASY ASIC, i.e., it replicates the fixed-point arithmetic of the hardware datapath. In what follows, we call this golden model \emph{HASY model}. 
\fref{fig:eval_waveform_grid} depicts alias-free hard-sync sine and sawtooth waveforms generated by the HASY model.
In addition, \fref{fig:sand_heart} shows mirrored sync and pulsar sync applied to unconventional waveforms.

For the evaluation, we compare the output signals generated by the HASY model against the following two baselines:
\begin{itemize}
\item The first baseline (called ``float'') is a floating-point reference software implementation that executes the spectral-resampling transform and additive synthesis as described in \fref{sec:algorithm}, with up to  $512$ harmonics.
\item The second baseline (called ``analytical'') starts from the ideal hard-sync signal that is not bandlimited. We then analytically compute the Fourier-series coefficients and resynthesize the time-domain signal with up to $512$ harmonics.
\end{itemize}

We generate \qty{1}{\s} of output samples (corresponding to $L=96\,000$ samples) with the HASY model and the two baselines for different following-oscillator waveforms and different values of $P$ (leading-to-following-oscillator period ratio).
Then, we compute the signal-to-noise and distortion ratio (\SINAD) in the sample domain as follows:
\begin{align}
    \SINAD &= 10 \log_{10}\! \left( \frac{ \sum_{\ell=1}^L|\achard{s}[\ell]|^2 }{ \sum_{\ell=1}^L |\achard{s}[\ell]-\tilde{s}[\ell]|^2 } \right)\!.
\end{align}
Here, $\tilde{s}[\cdot]$ represents the reference signal (i.e., the floating-point reference or the analytical reference).

\subsection{SINAD Comparison}

\begin{figure}[tp]
\centering
\includegraphics[width=0.95\columnwidth]{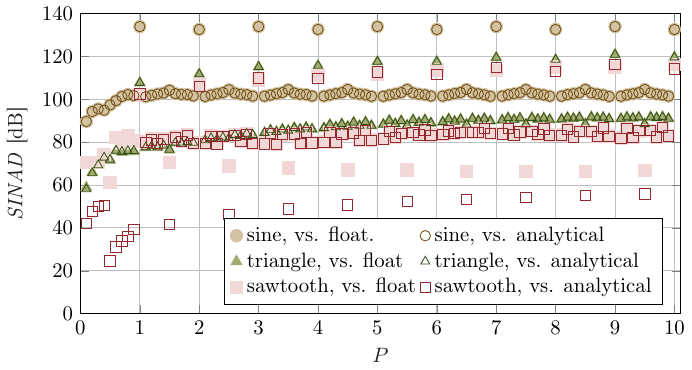}
\caption{\SINAD versus $P$ for hard sync of three different following-oscillator waveforms at $\flead=\qty{94}{\Hz}$.}
\label{fig:sinad_vs_p}
\end{figure}

\fref{fig:sinad_vs_p} shows the \SINAD for three different following-oscillator waveforms (sine,  triangle, and sawtooth) for different values of $P$. The leading-oscillator frequency is kept at $\flead = \qty{94}{\Hz}$, which corresponds to $510$ harmonics below $\fs/2 = \qty{48}{\kHz}$, so almost all of the harmonics available to HASY are utilized.
From the results in \fref{fig:sinad_vs_p}, we make the following observations:

If $P$ is an integer, then the \SINAD is high, and our approach yields high accuracy. 
This is not unexpected as for $P \in \mathbb{N}$, the spectral-resampling transform corresponds to a simple Fourier-series coefficient remapping, where the $n$th harmonic is mapped to the $nP$th harmonic.
If $P < 1$, then the \SINAD drops.
This performance drop is because our method starts from a following-oscillator waveform represented by only $512$ Fourier-series coefficients, i.e., already bandlimited (cf. Remark \ref{rm:bandlimited-following-osc}).
When performing the spectral-resampling transform for $P < 1$, the hard-sync Fourier-series coefficients largely depend on the higher-index Fourier-series coefficients of the following-oscillator waveform. 
However, since $N$ is fixed to $512$, these higher-index Fourier-series coefficients might not be available for resampling.
Generally, this effect worsens for smaller values of $P$ and $\flead$.

\begin{remark}
This observation reveals a key limitation of our approach: Resampling can only transform harmonics that are already present in the Fourier-series coefficients of the free-running following waveform.
\end{remark}

For higher $\flead$ (not shown in \fref{fig:sinad_vs_p}), the performance drop at small $P$ is less pronounced, since the bandlimited following-oscillator constraint mainly affects higher harmonics (which are no longer within the band of interest at high~$\flead$).

\section{Conclusions}

We have introduced a method for alias-free oscillator synchronization based on additive synthesis, which supports three synchronization modes and arbitrary bandlimited following-oscillator waveforms. 
To address the high computational cost of our approach, we have developed an ASIC, which implements the proposed method with a latency of fewer than five audio samples, demonstrating practical, real-time realization of our approach. 
A comparison of the ASIC's output with both an analytical (ideal) and a floating-point baseline reveals that our implementation achieves \SINAD values above \qty{41}{\dB} (if the following oscillator's frequency exceeds that of the leading oscillator).

There are several avenues for future work.
First, we are working on a polyphonic version of the HASY ASIC that also fixes its implementation bugs.
Second, for certain waveforms, Fourier-series coefficients could be derived directly from the analytical synchronized functions.
Finally, we are developing a Eurorack module to demonstrate the HASY ASIC in a practical instrument.

On the companion page of this article we provide a reference Python (NumPy) implementation and generated audio examples.\footnote{\url{https://github.com/IIP-Group/hasy-python}}
\vspace{-1mm}
\section{Acknowledgments}
\vspace{-1mm}
The authors thank the Microelectronics Design Center of ETH Zurich for their help with the tape-out of the HASY ASIC.


\bibliographystyle{bib/IEEEtranDAFx}
\bibliography{bib/VIPabbrv, bib/confs-jrnls, bib/26DAFx_HardSync}


\end{document}

%% file: macros/vmr-symbols-vecbold.tex
\usepackage{amssymb}
\usepackage{amsfonts}
\usepackage{mathrsfs}
\usepackage{xspace}
\usepackage{bm}
\usepackage{upgreek}

\newcommand{\safemath}[2]{\newcommand{#1}{\ensuremath{#2}\xspace}}

\safemath{\bma}{\mathbf{a}}
\safemath{\bmb}{\mathbf{b}}
\safemath{\bmc}{\mathbf{c}}
\safemath{\bmd}{\mathbf{d}}
\safemath{\bme}{\mathbf{e}}
\safemath{\bmf}{\mathbf{f}}
\safemath{\bmg}{\mathbf{g}}
\safemath{\bmh}{\mathbf{h}}
\safemath{\bmi}{\mathbf{i}}
\safemath{\bmj}{\mathbf{j}}
\safemath{\bmk}{\mathbf{k}}
\safemath{\bml}{\mathbf{l}}
\safemath{\bmm}{\mathbf{m}}
\safemath{\bmn}{\mathbf{n}}
\safemath{\bmo}{\mathbf{o}}
\safemath{\bmp}{\mathbf{p}}
\safemath{\bmq}{\mathbf{q}}
\safemath{\bmr}{\mathbf{r}}
\safemath{\bms}{\mathbf{s}}
\safemath{\bmt}{\mathbf{t}}
\safemath{\bmu}{\mathbf{u}}
\safemath{\bmv}{\mathbf{v}}
\safemath{\bmw}{\mathbf{w}}
\safemath{\bmx}{\mathbf{x}}
\safemath{\bmy}{\mathbf{y}}
\safemath{\bmz}{\mathbf{z}}
\safemath{\bmzero}{\mathbf{0}}
\safemath{\bmone}{\mathbf{1}}

\bmdefine{\biad}{a}
\bmdefine{\bibd}{b}
\bmdefine{\bicd}{c}
\bmdefine{\bidd}{d}
\bmdefine{\bied}{e}
\bmdefine{\bifd}{f}
\bmdefine{\bigd}{g}
\bmdefine{\bihd}{h}
\bmdefine{\biid}{i}
\bmdefine{\bijd}{j}
\bmdefine{\bikd}{k}
\bmdefine{\bild}{l}
\bmdefine{\bimd}{m}
\bmdefine{\bind}{n}
\bmdefine{\biod}{o}
\bmdefine{\bipd}{p}
\bmdefine{\biqd}{q}
\bmdefine{\bird}{r}
\bmdefine{\bisd}{s}
\bmdefine{\bitd}{t}
\bmdefine{\biud}{u}
\bmdefine{\bivd}{v}
\bmdefine{\biwd}{w}
\bmdefine{\bixd}{x}
\bmdefine{\biyd}{y}
\bmdefine{\bizd}{z}

\bmdefine{\bixid}{\xi}
\bmdefine{\bilambdad}{\lambda}
\bmdefine{\bimud}{\mu}
\bmdefine{\bithetad}{\theta}
\bmdefine{\biphid}{\phi}
\bmdefine{\bideltad}{\delta}

\safemath{\bmia}{\biad}
\safemath{\bmib}{\bibd}
\safemath{\bmic}{\bicd}
\safemath{\bmid}{\bidd}
\safemath{\bmie}{\bied}
\safemath{\bmif}{\bifd}
\safemath{\bmig}{\bigd}
\safemath{\bmih}{\bihd}
\safemath{\bmii}{\biid}
\safemath{\bmij}{\bijd}
\safemath{\bmik}{\bikd}
\safemath{\bmil}{\bild}
\safemath{\bmim}{\bimd}
\safemath{\bmin}{\bind}
\safemath{\bmio}{\biod}
\safemath{\bmip}{\bipd}
\safemath{\bmiq}{\biqd}
\safemath{\bmir}{\bird}
\safemath{\bmis}{\bisd}
\safemath{\bmit}{\bitd}
\safemath{\bmiu}{\biud}
\safemath{\bmiv}{\bivd}
\safemath{\bmiw}{\biwd}
\safemath{\bmix}{\bixd}
\safemath{\bmiy}{\biyd}
\safemath{\bmiz}{\bizd}

\safemath{\bmxi}{\bixid}
\safemath{\bmlambda}{\bilambdad}
\safemath{\bmmu}{\bimud}
\safemath{\bmtheta}{\bithetad}
\safemath{\bmphi}{\biphid}
\safemath{\bmdelta}{\bideltad}

\safemath{\bA}{\mathbf{A}}
\safemath{\bB}{\mathbf{B}}
\safemath{\bC}{\mathbf{C}}
\safemath{\bD}{\mathbf{D}}
\safemath{\bE}{\mathbf{E}}
\safemath{\bF}{\mathbf{F}}
\safemath{\bG}{\mathbf{G}}
\safemath{\bH}{\mathbf{H}}
\safemath{\bI}{\mathbf{I}}
\safemath{\bJ}{\mathbf{J}}
\safemath{\bK}{\mathbf{K}}
\safemath{\bL}{\mathbf{L}}
\safemath{\bM}{\mathbf{M}}
\safemath{\bN}{\mathbf{N}}
\safemath{\bO}{\mathbf{O}}
\safemath{\bP}{\mathbf{P}}
\safemath{\bQ}{\mathbf{Q}}
\safemath{\bR}{\mathbf{R}}
\safemath{\bS}{\mathbf{S}}
\safemath{\bT}{\mathbf{T}}
\safemath{\bU}{\mathbf{U}}
\safemath{\bV}{\mathbf{V}}
\safemath{\bW}{\mathbf{W}}
\safemath{\bX}{\mathbf{X}}
\safemath{\bY}{\mathbf{Y}}
\safemath{\bZ}{\mathbf{Z}}

\safemath{\bZero}{\mathbf{0}}
\safemath{\bOne}{\mathbf{1}}
\safemath{\bDelta}{\mathbf{\Delta}}
\safemath{\bLambda}{\mathbf{\UpLambda}}
\safemath{\bPhi}{\mathbf{\Upphi}}
\safemath{\bSigma}{\mathbf{\Upsigma}}
\safemath{\bOmega}{\mathbf{\Upomega}}
\safemath{\bTheta}{\mathbf{\Uptheta}}

\bmdefine{\biAd}{A}
\bmdefine{\biBd}{B}
\bmdefine{\biCd}{C}
\bmdefine{\biDd}{D}
\bmdefine{\biEd}{E}
\bmdefine{\biFd}{F}
\bmdefine{\biGd}{G}
\bmdefine{\biHd}{H}
\bmdefine{\biId}{I}
\bmdefine{\biJd}{J}
\bmdefine{\biKd}{K}
\bmdefine{\biLd}{L}
\bmdefine{\biMd}{M}
\bmdefine{\biOd}{N}
\bmdefine{\biPd}{O}
\bmdefine{\biQd}{P}
\bmdefine{\biRd}{R}
\bmdefine{\biSd}{S}
\bmdefine{\biTd}{T}
\bmdefine{\biUd}{U}
\bmdefine{\biVd}{V}
\bmdefine{\biWd}{W}
\bmdefine{\biXd}{X}
\bmdefine{\biYd}{Y}
\bmdefine{\biZd}{Z}

\bmdefine{\biDelta}{\Delta}
\bmdefine{\biLambda}{\Lambda}
\bmdefine{\biPhi}{\Phi}
\bmdefine{\biSigma}{\Sigma}
\bmdefine{\biOmega}{\Omega}
\bmdefine{\biTheta}{\Theta}

\safemath{\bimA}{\biAd}
\safemath{\bimB}{\biBd}
\safemath{\bimC}{\biCd}
\safemath{\bimD}{\biDd}
\safemath{\bimE}{\biEd}
\safemath{\bimF}{\biFd}
\safemath{\bimG}{\biGd}
\safemath{\bimH}{\biHd}
\safemath{\bimI}{\biId}
\safemath{\bimJ}{\biJd}
\safemath{\bimK}{\biKd}
\safemath{\bimL}{\biLd}
\safemath{\bimM}{\biMd}
\safemath{\bimN}{\biNd}
\safemath{\bimO}{\biOd}
\safemath{\bimP}{\biPd}
\safemath{\bimQ}{\biQd}
\safemath{\bimR}{\biRd}
\safemath{\bimS}{\biSd}
\safemath{\bimT}{\biTd}
\safemath{\bimU}{\biUd}
\safemath{\bimV}{\biVd}
\safemath{\bimW}{\biWd}
\safemath{\bimX}{\biXd}
\safemath{\bimY}{\biYd}
\safemath{\bimZ}{\biZd}

\safemath{\bimDelta}{\biDelta}
\safemath{\bimLambda}{\biLambda}
\safemath{\bimPhi}{\biPhi}
\safemath{\bimSigma}{\biSigma}
\safemath{\bimOmega}{\biOmega}
\safemath{\bimTheta}{\biTheta}

\safemath{\setA}{\mathcal{A}}
\safemath{\setB}{\mathcal{B}}
\safemath{\setC}{\mathcal{C}}
\safemath{\setD}{\mathcal{D}}
\safemath{\setE}{\mathcal{E}}
\safemath{\setF}{\mathcal{F}}
\safemath{\setG}{\mathcal{G}}
\safemath{\setH}{\mathcal{H}}
\safemath{\setI}{\mathcal{I}}
\safemath{\setJ}{\mathcal{J}}
\safemath{\setK}{\mathcal{K}}
\safemath{\setL}{\mathcal{L}}
\safemath{\setM}{\mathcal{M}}
\safemath{\setN}{\mathcal{N}}
\safemath{\setO}{\mathcal{O}}
\safemath{\setP}{\mathcal{P}}
\safemath{\setQ}{\mathcal{Q}}
\safemath{\setR}{\mathcal{R}}
\safemath{\setS}{\mathcal{S}}
\safemath{\setT}{\mathcal{T}}
\safemath{\setU}{\mathcal{U}}
\safemath{\setV}{\mathcal{V}}
\safemath{\setW}{\mathcal{W}}
\safemath{\setX}{\mathcal{X}}
\safemath{\setY}{\mathcal{Y}}
\safemath{\setZ}{\mathcal{Z}}
\safemath{\emptySet}{\varnothing}

\safemath{\colA}{\mathscr{A}}
\safemath{\colB}{\mathscr{B}}
\safemath{\colC}{\mathscr{C}}
\safemath{\colD}{\mathscr{D}}
\safemath{\colE}{\mathscr{E}}
\safemath{\colF}{\mathscr{F}}
\safemath{\colG}{\mathscr{G}}
\safemath{\colH}{\mathscr{H}}
\safemath{\colI}{\mathscr{I}}
\safemath{\colJ}{\mathscr{J}}
\safemath{\colK}{\mathscr{K}}
\safemath{\colL}{\mathscr{L}}
\safemath{\colM}{\mathscr{M}}
\safemath{\colN}{\mathscr{N}}
\safemath{\colO}{\mathscr{O}}
\safemath{\colP}{\mathscr{P}}
\safemath{\colQ}{\mathscr{Q}}
\safemath{\colR}{\mathscr{R}}
\safemath{\colS}{\mathscr{S}}
\safemath{\colT}{\mathscr{T}}
\safemath{\colU}{\mathscr{U}}
\safemath{\colV}{\mathscr{V}}
\safemath{\colW}{\mathscr{W}}
\safemath{\colX}{\mathscr{X}}
\safemath{\colY}{\mathscr{Y}}
\safemath{\colZ}{\mathscr{Z}}

\safemath{\opA}{\mathbb{A}}
\safemath{\opB}{\mathbb{B}}
\safemath{\opC}{\mathbb{C}}
\safemath{\opD}{\mathbb{D}}
\safemath{\opE}{\mathbb{E}}
\safemath{\opF}{\mathbb{F}}
\safemath{\opG}{\mathbb{G}}
\safemath{\opH}{\mathbb{H}}
\safemath{\opI}{\mathbb{I}}
\safemath{\opJ}{\mathbb{J}}
\safemath{\opK}{\mathbb{K}}
\safemath{\opL}{\mathbb{L}}
\safemath{\opM}{\mathbb{M}}
\safemath{\opN}{\mathbb{N}}
\safemath{\opO}{\mathbb{O}}
\safemath{\opP}{\mathbb{P}}
\safemath{\opQ}{\mathbb{Q}}
\safemath{\opR}{\mathbb{R}}
\safemath{\opS}{\mathbb{S}}
\safemath{\opT}{\mathbb{T}}
\safemath{\opU}{\mathbb{U}}
\safemath{\opV}{\mathbb{V}}
\safemath{\opW}{\mathbb{W}}
\safemath{\opX}{\mathbb{X}}
\safemath{\opY}{\mathbb{Y}}
\safemath{\opZ}{\mathbb{Z}}
\safemath{\opZero}{\mathbb{O}}
\safemath{\identityop}{\opI}

\safemath{\veca}{\bma}
\safemath{\vecb}{\bmb}
\safemath{\vecc}{\bmc}
\safemath{\vecd}{\bmd}
\safemath{\vece}{\bme}
\safemath{\vecf}{\bmf}
\safemath{\vecg}{\bmg}
\safemath{\vech}{\bmh}
\safemath{\veci}{\bmi}
\safemath{\vecj}{\bmj}
\safemath{\veck}{\bmk}
\safemath{\vecl}{\bml}
\safemath{\vecm}{\bmm}
\safemath{\vecn}{\bmn}
\safemath{\veco}{\bmo}
\safemath{\vecp}{\bmp}
\safemath{\vecq}{\bmq}
\safemath{\vecr}{\bmr}
\safemath{\vecs}{\bms}
\safemath{\vect}{\bmt}
\safemath{\vecu}{\bmu}
\safemath{\vecv}{\bmv}
\safemath{\vecw}{\bmw}
\safemath{\vecx}{\bmx}
\safemath{\vecy}{\bmy}
\safemath{\vecz}{\bmz}

\safemath{\veczero}{\bmzero}
\safemath{\vecone}{\bmone}
\safemath{\vecxi}{\bmxi}
\safemath{\veclambda}{\bmlambda}
\safemath{\vecmu}{\bmmu}
\safemath{\vectheta}{\bmtheta}
\safemath{\vecphi}{\bmphi}
\safemath{\vecdelta}{\bmdelta}

\safemath{\matA}{\bA}
\safemath{\matB}{\bB}
\safemath{\matC}{\bC}
\safemath{\matD}{\bD}
\safemath{\matE}{\bE}
\safemath{\matF}{\bF}
\safemath{\matG}{\bG}
\safemath{\matH}{\bH}
\safemath{\matI}{\bI}
\safemath{\matJ}{\bJ}
\safemath{\matK}{\bK}
\safemath{\matL}{\bL}
\safemath{\matM}{\bM}
\safemath{\matN}{\bN}
\safemath{\matO}{\bO}
\safemath{\matP}{\bP}
\safemath{\matQ}{\bQ}
\safemath{\matR}{\bR}
\safemath{\matS}{\bS}
\safemath{\matT}{\bT}
\safemath{\matU}{\bU}
\safemath{\matV}{\bV}
\safemath{\matW}{\bW}
\safemath{\matX}{\bX}
\safemath{\matY}{\bY}
\safemath{\matZ}{\bZ}
\safemath{\matzero}{\bmzero}

\safemath{\matDelta}{\bDelta}
\safemath{\matLambda}{\bLambda}
\safemath{\matPhi}{\bPhi}
\safemath{\matSigma}{\bSigma}
\safemath{\matOmega}{\bOmega}
\safemath{\matTheta}{\bTheta}

\safemath{\matidentity}{\matI}
\safemath{\matone}{\matO}

\safemath{\rnda}{A}
\safemath{\rndb}{B}
\safemath{\rndc}{C}
\safemath{\rndd}{D}
\safemath{\rnde}{E}
\safemath{\rndf}{F}
\safemath{\rndg}{G}
\safemath{\rndh}{H}
\safemath{\rndi}{I}
\safemath{\rndj}{J}
\safemath{\rndk}{K}
\safemath{\rndl}{L}
\safemath{\rndm}{M}
\safemath{\rndn}{N}
\safemath{\rndo}{O}
\safemath{\rndp}{P}
\safemath{\rndq}{Q}
\safemath{\rndr}{R}
\safemath{\rnds}{S}
\safemath{\rndt}{T}
\safemath{\rndu}{U}
\safemath{\rndv}{V}
\safemath{\rndw}{W}
\safemath{\rndx}{X}
\safemath{\rndy}{Y}
\safemath{\rndz}{Z}

\safemath{\rveca}{\bimA}
\safemath{\rvecb}{\bimB}
\safemath{\rvecc}{\bimC}
\safemath{\rvecd}{\bimD}
\safemath{\rvece}{\bimE}
\safemath{\rvecf}{\bimF}
\safemath{\rvecg}{\bimG}
\safemath{\rvech}{\bimH}
\safemath{\rveci}{\bimI}
\safemath{\rvecj}{\bimJ}
\safemath{\rveck}{\bimK}
\safemath{\rvecl}{\bimL}
\safemath{\rvecm}{\bimM}
\safemath{\rvecn}{\bimN}
\safemath{\rveco}{\bomO}
\safemath{\rvecp}{\bimP}
\safemath{\rvecq}{\bimQ}
\safemath{\rvecr}{\bimR}
\safemath{\rvecs}{\bimS}
\safemath{\rvect}{\bimT}
\safemath{\rvecu}{\bimU}
\safemath{\rvecv}{\bimV}
\safemath{\rvecw}{\bimW}
\safemath{\rvecx}{\bimX}
\safemath{\rvecy}{\bimY}
\safemath{\rvecz}{\bimZ}

\safemath{\rvecxi}{\bmxi}
\safemath{\rveclambda}{\bmlambda}
\safemath{\rvecmu}{\bmmu}
\safemath{\rvectheta}{\bmtheta}
\safemath{\rvecphi}{\bmphi}

\safemath{\rmatA}{\bimA}
\safemath{\rmatB}{\bimB}
\safemath{\rmatC}{\bimC}
\safemath{\rmatD}{\bimD}
\safemath{\rmatE}{\bimE}
\safemath{\rmatF}{\bimF}
\safemath{\rmatG}{\bimG}
\safemath{\rmatH}{\bimH}
\safemath{\rmatI}{\bimI}
\safemath{\rmatJ}{\bimJ}
\safemath{\rmatK}{\bimK}
\safemath{\rmatL}{\bimL}
\safemath{\rmatM}{\bimM}
\safemath{\rmatN}{\bimN}
\safemath{\rmatO}{\bimO}
\safemath{\rmatP}{\bimP}
\safemath{\rmatQ}{\bimQ}
\safemath{\rmatR}{\bimR}
\safemath{\rmatS}{\bimS}
\safemath{\rmatT}{\bimT}
\safemath{\rmatU}{\bimU}
\safemath{\rmatV}{\bimV}
\safemath{\rmatW}{\bimW}
\safemath{\rmatX}{\bimX}
\safemath{\rmatY}{\bimY}
\safemath{\rmatZ}{\bimZ}

\safemath{\rmatDelta}{\bimDelta}
\safemath{\rmatLambda}{\bimLambda}
\safemath{\rmatPhi}{\bimPhi}
\safemath{\rmatSigma}{\bimSigma}
\safemath{\rmatOmega}{\bimOmega}
\safemath{\rmatTheta}{\bimTheta}

%% file: macros/standard-macros.tex
\usepackage{amssymb}
\usepackage{amsfonts}
\usepackage{mathrsfs}
\usepackage{xspace}
\usepackage{bm}
\usepackage{fancyref}
\usepackage{textcomp}

\usepackage{multirow}
\usepackage{stmaryrd}

\newenvironment{textbmatrix}{	\setlength{\arraycolsep}{2.5pt}%
								\big[\begin{matrix}}{\end{matrix}\big]%
								\raisebox{0.08ex}{\vphantom{M}}}

\def\be{\begin{equation}}
\def\ee{\end{equation}}
\def\een{\nonumber \end{equation}}
\def\mat{\begin{bmatrix}}
\def\emat{\end{bmatrix}}
\def\btm{\begin{textbmatrix}}
\def\etm{\end{textbmatrix}}

\def\ba#1\ea{\begin{align}#1\end{align}}
\def\bas#1\eas{\begin{align*}#1\end{align*}}
\def\bs#1\es{\begin{split}#1\end{split}}
\def\bg#1\eg{\begin{gather}#1\end{gather}}
\def\bml#1\eml{\begin{multline}#1\end{multline}}
\def\bi#1\ei{\begin{itemize}#1\end{itemize}}

\DeclareMathOperator{\sinc}{sinc}			

\safemath{\dirac}{\delta}					
\safemath{\krond}{\dirac}					

\safemath{\upto}{\uparrow}
\safemath{\downto}{\downarrow}
\safemath{\iu}{j}							
\safemath{\ev}{\lambda}						
\safemath{\hilseqspace}{l^{2}}				
\newcommand{\banachfunspace}[1]{\setL^{#1}}	
\safemath{\hilfunspace}{\banachfunspace{2}}	

\safemath{\SNR}{\textit{SNR}} 				
\safemath{\PAR}{\textit{PAR}} 				
\safemath{\No}{N_0}							
\safemath{\Es}{E_s}							
\safemath{\Eb}{E_b}							
\safemath{\EbNo}{\frac{\Eb}{\No}}
\safemath{\EsNo}{\frac{\Es}{\No}}

\DeclareMathOperator{\CHop}{\ensuremath{\opH}} 
\safemath{\tvir}{\rndh_{\CHop}}				
\safemath{\tvtf}{\rndl_{\CHop}}				
\safemath{\spf}{\rnds_{\CHop}}				
\safemath{\bff}{H_{\CHop}}					

\safemath{\ircf}{r_{h}}						
\safemath{\tftvcf}{r_{s}}					
\safemath{\tfcf}{r_{l}}						
\safemath{\bfcf}{r_{H}}						

\safemath{\tcorr}{c_h}						
\safemath{\scf}{c_{s}}						
\safemath{\tfcorr}{c_{l}}					
\safemath{\fcorr}{c_{H}}						

\safemath{\mi}{I}							
\safemath{\capacity}{C}						

\safemath{\normal}{\mathcal{N}}			
\safemath{\jpg}{\mathcal{CN}}			
\safemath{\mchain}{\leftrightarrow}		

\safemath{\dB}{\,\mathrm{dB}}
\safemath{\dBm}{\,\mathrm{dBm}}
\safemath{\Hz}{\,\mathrm{Hz}}
\safemath{\kHz}{\,\mathrm{kHz}}
\safemath{\MHz}{\,\mathrm{MHz}}
\safemath{\GHz}{\,\mathrm{GHz}}
\safemath{\s}{\,\mathrm{s}}
\safemath{\ms}{\,\mathrm{ms}}
\safemath{\mus}{\,\mathrm{\text{\textmu}s}}
\safemath{\ns}{\,\mathrm{ns}}
\safemath{\ps}{\,\mathrm{ps}}
\safemath{\meter}{\,\mathrm{m}}
\safemath{\mm}{\,\mathrm{mm}}
\safemath{\cm}{\,\mathrm{cm}}
\safemath{\m}{\,\mathrm{m}}
\safemath{\W}{\,\mathrm{W}}
\safemath{\mW}{\, \mathrm{mW}}
\safemath{\J}{\,\mathrm{J}}
\safemath{\K}{\,\mathrm{K}}
\safemath{\bit}{\,\mathrm{bit}}
\safemath{\nat}{\,\mathrm{nat}}

\safemath{\define}{\triangleq}			

\safemath{\equivalent}{\sim}
\safemath{\distas}{\sim}					
\safemath{\sdiff}{\Delta}				

\safemath{\reals}{\mathbb{R}}
\safemath{\positivereals}{\reals_{+}}
\safemath{\integers}{\mathbb{Z}}
\safemath{\posint}{\integers_{+}}
\safemath{\naturals}{\mathbb{N}}
\safemath{\posnaturals}{\naturals_{+}}
\safemath{\complexset}{\mathbb{C}}
\safemath{\rationals}{\mathbb{Q}}

\newcommand*{\fancyrefapplabelprefix}{app}		
\newcommand*{\fancyrefthmlabelprefix}{thm}		
\newcommand*{\fancyreflemlabelprefix}{lem}		
\newcommand*{\fancyrefcorlabelprefix}{cor}		
\newcommand*{\fancyrefdeflabelprefix}{def}		
\newcommand*{\fancyrefproplabelprefix}{prop}		
\newcommand*{\fancyrefexmpllabelprefix}{exmpl}
\newcommand*{\fancyrefalglabelprefix}{alg}		
\newcommand*{\fancyreftbllabelprefix}{tbl}		

\frefformat{vario}{\fancyrefseclabelprefix}{Section~#1}
\frefformat{vario}{\fancyrefthmlabelprefix}{Theorem.~#1}
\frefformat{vario}{\fancyreftbllabelprefix}{Table~#1}
\frefformat{vario}{\fancyreflemlabelprefix}{Lemma~#1}
\frefformat{vario}{\fancyrefcorlabelprefix}{Corollary~#1}
\frefformat{vario}{\fancyrefdeflabelprefix}{Definition~#1}
\frefformat{vario}{\fancyreffiglabelprefix}{Figure~#1}
\frefformat{vario}{\fancyrefapplabelprefix}{Appendix~#1}
\frefformat{vario}{\fancyrefeqlabelprefix}{(#1)}
\frefformat{vario}{\fancyrefproplabelprefix}{Proposition~#1}
\frefformat{vario}{\fancyrefexmpllabelprefix}{Example~#1}
\frefformat{vario}{\fancyrefalglabelprefix}{Algorithm~#1}

%% file: macros/defs.tex
 \newtheorem{remark}{Remark}
 \newtheorem*{remark*}{Remark}

\safemath{\dictab}{[\,\dicta\,\,\dictb\,]}

\safemath{\ysig}{\bmy}
\safemath{\ysighat}{\hat{\ysig}}
\safemath{\ysigdim}{M}
\safemath{\xsig}{\bmx}
\safemath{\xsigdim}{N}
\safemath{\nx}{n_x}
\safemath{\zsig}{\bmz}
\safemath{\zsigdim}{\ysigdim}
\safemath{\rsig}{\bmr}
\safemath{\Adict}{\bA}
\safemath{\Adicttilde}{\widetilde{\Adict}}
\safemath{\Adictdim}{\outputdim\times\xsigdim}
\safemath{\avec}{\bma}
\safemath{\avectilde}{\tilde{\avec}}
\safemath{\Bdict}{\bB}
\safemath{\Bdicttilde}{\widetilde{\Bdict}}
\safemath{\Cdict}{\bC}
\safemath{\cvec}{\bmc}
\safemath{\Ddict}{\bD}
\safemath{\Ddictdim}{\ysigdim\times\xsigdim}
\safemath{\dvec}{\bmd}
\safemath{\Ddicttilde}{\widetilde{\bD}}
\safemath{\Bonb}{\bB}
\safemath{\bvec}{\bmb}
\safemath{\Bonbdim}{\ysigdim\times\ysigdim}
\safemath{\noise}{\bmn}
\safemath{\noisedim}{\ysigim}
\safemath{\err}{\bme}
\safemath{\errdim}{\ysigdim}
\safemath{\errset}{\setE}
\safemath{\nerr}{n_e}
\safemath{\delop}{\bP_\errset}
\safemath{\delopc}{\bP_{{\errset}^c}}

\safemath{\cplxi}{\imath}
\safemath{\cplxj}{\jmath}

\safemath{\dict}{\matD}
\safemath{\inputdim}{N}		
\safemath{\outputdim}{M}		
\safemath{\sparsity}{S}	
\safemath{\inputdimA}{{N_a}}	
\safemath{\inputdimB}{{N_b}}	
\safemath{\elemA}{{n_a}}	
\safemath{\elemB}{{n_b}}	
\safemath{\resA}{\matR_a}	
\safemath{\resB}{\matR_b}	
\safemath{\subD}{\matS} 
\safemath{\subA}{\matS_a} 
\safemath{\subB}{\matS_b} 
\safemath{\dicta}{\matA} 	
\safemath{\dictb}{\matB} 	
\safemath{\hollowS}{H}
\safemath{\hollowA}{H_a}
\safemath{\hollowB}{H_b}
\safemath{\cross}{Z}
\safemath{\coh}{\mu_d}			
\safemath{\coha}{\mu_a}			
\safemath{\cohb}{\mu_b}			
\safemath{\mubs}{\nu}	
\safemath{\cohm}{\mu_m} 
\safemath{\dictset}{\setD}	
\safemath{\dictsetp}{\dictset(\coh,\coha,\cohb)}	
\safemath{\dictsetgen}{\dictset_\text{gen}}
\safemath{\dictsetgenp}{\dictsetgen(\coh)}
\safemath{\dictsetonb}{\dictset_\text{onb}}
\safemath{\dictsetonbp}{\dictsetonb(\coh)}

\safemath{\leftside}{U}
\safemath{\rightsideA}{R_a}
\safemath{\rightsideB}{R_b}

\safemath{\indexS}{\setI_S} 

\safemath{\na}{n_a}			
\safemath{\nb}{n_b}			
\safemath{\coeffa}{p_i}	
\safemath{\coeffb}{q_j}	
\safemath{\seta}{\setP}		
\safemath{\setb}{\setQ}     
\safemath{\setw}{\setW}	
\safemath{\setz}{\setZ}	
\safemath{\cola}{\veca}		
\safemath{\colb}{\vecb}		
\safemath{\cold}{\vecd}		
\safemath{\inputvec}{\vecx} 	
\safemath{\error}{\vece}	
\safemath{\noiseout}{\vecz} 	
\safemath{\inputvecel}{x}
\safemath{\inputveca}{\vecx_a}
\safemath{\inputvecb}{\vecx_b}
\safemath{\outputvec}{\vecy}	
\safemath{\lambdamin}{\lambda_{\mathrm{min}}}

\safemath{\elltwo}{\ell_2}
\safemath{\ellone}{\ell_1}
\safemath{\ellzero}{\ell_0}
\safemath{\ellinf}{\ell_\infty}
\safemath{\ellinftilde}{\ell_{\widetilde\infty}}
\safemath{\licard}{Z(\coh,\coha,\cohb)}
\safemath{\xsol}{\hat{x}}
\safemath{\xbord}{x_b}		
\safemath{\xstat}{x_s}		
\safemath{\xstatLone}{\tilde{x}_s}
\safemath{\order}{\mathcal{O}} 
\safemath{\scales}{\Theta} 
\safemath{\ones}{\mathbf{1}} 
\safemath{\zeroes}{\mathbf{0}} 
\safemath{\thlone}{\kappa(\coh,\cohb)} 
\safemath{\constoneA}{\delta} 
\safemath{\constoneB}{\epsilon} 
\safemath{\nlarge}{L}				   
\safemath{\sumlarge}{S_\nlarge}
\safemath{\maxlarger}{P_\nlarge}	   
\safemath{\Pzero}{\textrm{P0}}	
\safemath{\Pone}{\textrm{P1}}
\safemath{\vecfir}{\vecw}			 
\safemath{\vecsec}{\vecz}
\safemath{\elvecfir}{w}              
\safemath{\elvecsec}{z}				 
\safemath{\nlargefir}{n}
\safemath{\normout}{\gamma}
\safemath{\auxfun}{h}
\safemath{\supp}{\textrm{supp}}

\safemath{\indexa}{\ell}
\safemath{\indexb}{r}
\safemath{\indexc}{i}
\safemath{\indexd}{j}

\safemath{\project}{P}